\documentclass[a4paper,11pt]{article}

\usepackage[top=1in, bottom=1in, left=1.1in, right=1.1in]{geometry} 

\usepackage[T1]{fontenc}
\usepackage[utf8]{inputenc}

\usepackage{microtype}

\usepackage{authblk}

\usepackage[sc]{mathpazo}

\usepackage{setspace}

\usepackage[rm]{titlesec}

\titlelabel{\thetitle.\enspace} 

\titleformat*{\section}{\Large\itshape} 
\titleformat*{\subsection}{\large\itshape} 
\titleformat*{\subsubsection}{\normalsize\itshape} 
\titleformat*{\paragraph}{\itshape} 

\usepackage[inline]{enumitem}
\setlist{noitemsep}

\usepackage[bottom, multiple]{footmisc}

\usepackage[labelfont=sc, labelsep=period, font=small, textfont=it]{caption} 

\usepackage[title, titletoc]{appendix} 

\usepackage{csquotes}

\usepackage{threeparttable}
\usepackage{booktabs}
\usepackage{tabularx}
\usepackage{colortbl}
\usepackage{makecell}

\usepackage{floatrow}
\usepackage{array}
\newcommand{\PreserveBackslash}[1]{\let\temp=\\#1\let\\=\temp}
\newcolumntype{C}[1]{>{\PreserveBackslash\centering}m{#1}}

\newcolumntype{s}{>{\hsize=.5\hsize}X}

\DeclareMathSizes{12}{12}{9}{9}

\usepackage{amsmath, amsthm, amsfonts}
\usepackage{mathtools}
\usepackage{bm}
\usepackage{nicefrac}
\usepackage{esvect}  

\newcommand{\wh}[1]{\widehat{#1}}

\theoremstyle{definition}

\theoremstyle{plain}

\numberwithin{equation}{section} 

\makeatletter
\newcommand*{\defeq}{\mathrel{\rlap{%
                     \raisebox{0.3ex}{$\m@th\cdot$}}%
                     \raisebox{-0.3ex}{$\m@th\cdot$}}%
                     =}
\makeatother

\makeatletter
\newcommand*{\invdefeq}{=\mathrel{\rlap{%
											  \raisebox{0.3ex}{$\m@th\cdot$}}%
                        \raisebox{-0.3ex}{$\m@th\cdot$}}%
                        }
\makeatother

\DeclareMathOperator*{\argmax}{arg\,max}

\newcommand{\E}{\mathbb{E}}
\newcommand{\R}{\mathbb{R}}

\newcommand{\co}{\textsc{co}\oldstylenums{2}}

\usepackage[table,x11names, dvipsnames]{xcolor}
\definecolor{mygreen}{RGB}{0,128,0} 

\usepackage{graphicx}

\usepackage{pgfplots}
\usepackage{pgfplotstable}

\pgfplotsset{width=0.65\textwidth, height=0.65*0.618\textwidth, compat=1.15}
\usepgfplotslibrary{fillbetween}

\usepackage{caption}
\usepackage{subcaption}

\usepackage[separate-uncertainty=true]{siunitx}

\DeclareSIUnit{\annum}{yr}
\DeclareSIUnit{\year}{yr}
\DeclareSIUnit{\years}{yr}
\DeclareSIUnit{\cent}{cent}
\DeclareSIUnit{\cents}{cents}
\DeclareSIUnit{\euro}{\texteuro}
\DeclareSIUnit{\euros}{\texteuro}
\DeclareSIUnit{\co}{\textsc{co}\oldstylenums{2}}
\DeclareSIUnit{\tco}{t_{\textsc{co}\oldstylenums{2}}}
\DeclareSIUnit{\euroco}{\texteuro/t\textsc{co}\oldstylenums{2}}
\DeclareSIUnit{\dollarco}{\textdollar/t\textsc{co}\oldstylenums{2}e}

\usepackage{eurosym}
\usepackage{icomma} 

\usepackage{hyperref}

\hypersetup{
    colorlinks,
    linkcolor={red!50!black},
    citecolor={black},
    urlcolor={blue!60!black}}

\usepackage{apacite}

\let\cite\shortcite 
\let\citeA\shortciteA 
\let\citeNP\shortciteNP 
\let\citeauthor\shortciteauthor 

\AtBeginDocument{\urlstyle{APACsame}}

\newcommand{\doi}[1]{\url{https://doi.org/#1}}

\AtBeginDocument{}

\title{
    Likelihood-Based Ergodicity Transformations in Time Series Analysis\thanks{The author gratefully acknowledges the financial support of the Research Training Group 2153, "Energy Status Data---Informatics Methods for its Collection, Analysis and Exploitation", of the German Research Foundation (DFG). The author also thanks the participants of the Ergodicity Economics Conference 2023 for helpful discussions.}
}

\author{Anthony Britto\thanks{\href{mailto:anthony.britto@kit.edu}{anthony.britto@kit.edu}.}}

\affil{\normalsize Institute for Industrial Production, Karlsruhe Institute of Technology\protect\\ Hertzstr.~16 -- Building 06.33, 76187 Karlsruhe, Germany}

\begin{document}

\maketitle

\onehalfspacing

\begin{abstract}
	\noindent Time series often exhibit non-ergodic behaviour that complicates forecasting and inference. This article proposes a likelihood-based approach for estimating ergodicity transformations that addresses such challenges. The method is broadly compatible with standard models, including Gaussian processes, ARMA, and GARCH. A detailed simulation study using geometric and arithmetic Brownian motion demonstrates the ability of the approach to recover known ergodicity transformations. A further case study on the large macroeconomic database FRED-QD shows that incorporating ergodicity transformations can provide meaningful improvements over conventional transformations or naive specifications in applied work.

    \medskip
	
	\noindent \emph{Key words}: Ergodicity transformations, time series analysis, likelihood estimation, forecasting, Box-Cox transformation, FRED-QD.
	
	\medskip
	
	\noindent \emph{JEL classification}: C22; C18; C38.
\end{abstract}

\newpage

\section{Introduction}
\label{sec:intro}

In the study of empirical time series, it is common to have access to only a single realisation of length $T$ of the stochastic process of interest $\{ X_t \}_{t=1}^\infty$, observed as $\{ x_t^{(1)} \}_{t=1}^T$. In such a situation, only \emph{time-average} quantities can be computed, since \emph{ensemble-average} quantities require multiple independent realisations. For instance, the sample mean,
\begin{equation}
    \overline{x} = \frac{1}{T} \sum_{t = 1}^T x_t^{(1)}
\end{equation}
is a time average, whereas the ensemble-average mean
\begin{equation}
    \E[X_t] = \lim_{I \to \infty} \frac{1}{I} \sum_{i = 1}^I x_t^{(i)}\ ,
\end{equation}
where $\{ x_t^{(i)} \}$ is the $i$th independent realisation of the process at time $t$, cannot be estimated since additional realisations are unavailable. Nonetheless, it is ensemble-average quantities that are of primary interest, as they characterise the probabilistic structure of $X_t$, for instance, its distribution, mean function, autocovariance, or higher-order moments. The concept which bridges the gap between time and ensemble averages is \emph{ergodicity}. To pursue the above example, a covariance-stationary process is said to be \emph{ergodic for the mean} if the time-average mean $\overline{x}$ converges in probability to the ensemble-average mean $\E[X_t]$ as $T \to \infty$; similar definitions exist for ergodicity of higher moments \cite[Ch.~3.1]{Hamilton1994}. The property of ergodicity is hence of fundamental importance in the theory and practice of time series analysis, since it allows for the consistent inference of the statistical properties of $X_t$ from a single realisation.

Many stochastic processes of practical interest do not satisfy the ergodicity property, exhibiting instead a phenomenon known as \emph{ergodicity breaking} \cite{Peters2013, Bel2005}. In such cases, time averages computed from a single trajectory do not converge to ensemble averages even as the observation window lengthens indefinitely. However, in certain instances, such non‐ergodic processes can be \emph{made} ergodic through the use of suitable transformations \cite{Magdziarz2020, Magdziarz2019}. The canonical example is geometric Brownian motion, which we write in discrete time with step $\Delta t$ as
\begin{equation}
    X_t = X_0\,\exp{\left( \left( \mu - \tfrac{1}{2}\sigma^2 \right) t + \sigma\,W_t\right)}\ ,
\end{equation}
where $X_0 > 0$ is the initial condition, $\mu \in \R$ is the drift, $\sigma > 0$ is the volatility, and $W_t$ satisfies
\begin{equation}
    W_0 = 0\ ,\quad \Delta W_t \defeq W_t - W_{t-\Delta t} \overset{\text{i.i.d.}}{\sim} \mathcal{N}(0,\Delta t)\ .
\label{eq:wiener}
\end{equation}
A natural goal is the consistent inference of $\mu$ and $\sigma$, which fully characterise $X_t$. However, since $X_t$ is non-ergodic, time-average quantities of a single realisation, such as a simple average of percentage changes, are path-dependent, and do not converge to the ensemble averages implied by $\mu$ and $\sigma$ \cite{Peters2013}. The remedy is to examine instead the transformed process
\begin{equation}
    \Delta \log{X_t} = \left( \mu - \tfrac{1}{2}\sigma^2 \right) \Delta t + \sigma \Delta W_t\ ,
\end{equation}
which follows an i.i.d. Gaussian law and is hence ergodic for all moments \cite[Ch.~3.1]{Hamilton1994}. The consistent estimation of $\mu$ and $\sigma$ thus becomes possible from a single, sufficiently long realisation of $X_t$ by studying the transformed process $\Delta \log{X_t}$; the $\Delta \log$ function may therefore be termed the \emph{ergodicity transformation} of geometric Brownian motion.\footnote{We use the term \enquote{ergodicity transformation} to refer to a change of variables that yields a stationary process whose time averages converge to ensemble averages in the sense used in time series analysis; it is not to be confused with an \enquote{ergodic transformation} in the formal sense of ergodic theory \cite{Walters2000}.}

In applications where the true underlying data-generating process is unknown, or a closed-form ergodicity transformation is unavailable, it becomes valuable to identify approximate ergodicity transformations. In addition to enabling consistent statistical inference, such transformations provide fundamental insight into the process's growth behaviour. In the case of geometric Brownian motion, for instance, recognizing the $\Delta \log$ as the ergodicity transformation identifies the process's growth as exponential, driven by the compounding of i.i.d.~Gaussian increments. In general, understanding the type of growth behaviour can be crucial for testing hypotheses, uncovering mechanisms driving the data-generating process, and improving model reliability in economics, finance, or climate science, where insights into long-term trends and fluctuations are essential \cite{Modis2013, Caporale2012, Chapman2013}.

This article proposes a simple likelihood procedure for this purpose, drawing on the seminal work of \citeA{Box1964} on the analysis of transformations. Although it is common in time series analysis to apply variance-stabilizing transformations such as the logarithm prior to fitting standard time series models (cf.~\citeNP{Hyndman2018}), there has to date been no study specifically targeting transformations resulting in approximate ergodicity.

The remainder of the article proceeds as follows. Section~\ref{sec:method} details the proposed likelihood approach to estimating ergodicity transformations. Section~\ref{sec:case_study_gbm} documents a simulation study of the proposed method using geometric and arithmetic Brownian motion. This is followed in Section~\ref{sec:case_study_fred} by an application of the proposed method to a factor analysis of a large macroeconomic database. Section~\ref{sec:discuss} concludes.

\section{Statistical ergodicity transformations}
\label{sec:method}

This section describes a likelihood approach for estimating ergodicity transformations, grounded in established results in time series analysis. The central idea is to transform the given time series so that an ergodic time-series model can be fitted. Standard and widely used examples of such models, which offer considerable flexibility, include Gaussian processes, autoregressive moving average (ARMA) models, and generalised autoregressive conditional heteroskedasticity (GARCH) models. To illustrate the approach, we first present the construction in the case of a Gaussian process, before extending the discussion to the ARMA and GARCH settings.

We use \citeA[Ch.~3.1]{Hamilton1994} as a reference for the following definitions and results. Recall that a stochastic process $\{ X_t \}_{t=1}^\infty$ is said to be \emph{Gaussian} if and only if for every finite set of indices $\{t_1, t_2, \ldots t_k\}$, the random variable
\begin{equation}
    \bm{X}_{t_1, t_2, \ldots, t_k} = (X_{t_1}, X_{t_2}, \ldots, X_{t_k})
\end{equation}
is multivariate Gaussian. A Gaussian process is said to be \emph{stationary} if neither the mean nor the autocovariances depend on the time $t$, namely, if
\begin{equation}
    \begin{cases}
        \E[X_t] = \mu\ , &\text{for all $t$}\ ,\\
        \E[(X_t - \mu)(X_{t-j} - \mu)] = \gamma_j\ , &\text{for all $t$ and any $j$}\ ,
    \end{cases}
\end{equation}
where $\mu \in \R$ is the mean level of the process, and $\gamma_j$ is the autocovariance function of $X_t$ at lag $j$.\footnote{This is the condition for so-called \emph{covariance-stationarity}, which also guarantees \emph{strict stationarity} in the case of a Gaussian process; see \cite[Ch.~3.1]{Hamilton1994}.} It is a fundamental result that a stationary Gaussian process with \emph{short-memory}, that is, with absolutely summable covariances
\begin{equation}
    \sum_{j = 0}^\infty | \gamma_j | < \infty\ ,
\end{equation}
is ergodic for all moments. Hence, any transformation that renders a given time series sufficiently Gaussian, stationary, and short-memory may be termed a statistical ergodicity transformation.

For ease of exposition, and motivated by the example of geometric Brownian motion, the following discussion first considers time series which are generated by the accumulation of random increments. It is often these generative increments, rather than the accumulated levels, that exhibit stable statistical properties; indeed, this principle undergirds the study of unit-root processes, which are ubiquitous in time-series analysis \cite[Ch.~15]{Hamilton1994}. So suppose that we are given a realisation $\bm{x} = \{ x_1, x_2, \ldots, x_T \}$ of a stochastic process $X_t$, and assume first that there exists some transformation $F$ such that the increment process $\Delta F(X_t)$ is \emph{Gaussian}. If we restrict to the one-parameter family of power transformations introduced in \cite{Box1964},
\begin{equation}
	F(x_t; \lambda) = \begin{cases}
		(x_t^\lambda - 1)/(\wh{\vartheta}^{\lambda-1} \lambda) & \lambda \neq 0\ ,\\
		\wh{\vartheta} \log{x_t} & \lambda = 0\ ,
	\end{cases}
\label{eq:boxcox}
\end{equation}
where $\wh{\vartheta}$ is the geometric mean of the observations, the problem simplifies to that of identifying $\wh{\lambda}$ such that
\begin{equation}
    \Delta F(\bm{x}; \wh{\lambda}) \sim \mathcal{N}_{T-1}(\bm{\mu}, \bm{\Sigma} )\ ,
\end{equation}
where $\mathcal{N}_{T-1}$ is a $T-1$-dimensional multivariate Gaussian distribution with mean vector $\bm{\mu} \in \R^{T-1}$ and covariance matrix $\bm{\Sigma} \in \R^{(T-1)\times(T-1)}$.\footnote{We abuse notation slightly by using $\Delta F(\bm{x}; \lambda)$ to denote the vector of the transformed realisation $\{ \Delta F(x_t; \lambda) \}_{t = 2}^T$. We have also employed here the marginalisation property of the multivariate Gaussian: if the entire set of increments is multivariate Gaussian, so is every subset of increments, which is precisely the condition needed for the increment process to be Gaussian.} If we include now the assumption that the increments are \emph{stationary}, the mean vector and covariance matrix in fact take the form
\begin{equation}
    \bm{\mu} = \mu \bm{1}\ , \quad \text{and} \quad  \bm{\Sigma}_{ij} = \gamma_{|i - j|}\ ,
\label{eq:mu_sigma_simplify}
\end{equation}
where $\bm{1}$ is the identity matrix and $\mu \in \R$ is a constant \cite{Xiao2012}. And if we introduce finally the assumption of \emph{short-memory}, it follows that for fixed $\lambda$, the maximised log likelihood up to a constant is given by
\begin{equation}
    L_{\text{max}}(\lambda) \defeq -\frac{1}{2} \left( \log{\det {\wh{\bm{\Sigma}}(\lambda)}} + ( \Delta F(\bm{x}; \lambda)- \wh{\mu}(\lambda)\bm{1})^\top \wh{\bm{\Sigma}}(\lambda)^{-1} ( \Delta F(\bm{x}; \lambda) - \wh{\mu}(\lambda)\bm{1}) \right)\ ,
\label{eq:llf_max}
\end{equation}
where $\wh{\bm{\Sigma}}(\lambda)$ is a consistent estimator of the autocovariance matrix, e.g.~as proposed in \citeA{McMurry2010}, and
\begin{equation}
    \wh{\mu}(\lambda) \defeq \frac{\bm{1}^\top \wh{\bm{\Sigma}}(\lambda)^{-1} \Delta F(\bm{x}; \lambda)}{\bm{1}^\top \wh{\bm{\Sigma}}(\lambda)^{-1} \bm{1}}
\end{equation}
is the maximised likelihood estimate of $\mu$ \cite{Xiao2012}.\footnote{The Jacobian of the transformation cancels in \eqref{eq:llf_max} due to the normalisation by the geometric mean, as in \citeA{Box1964}. This normalisation also aids the optimisation over $\lambda$ by making the problem scale-free.} The desired estimate of the ergodicity transformation parameter is hence given by
\begin{equation}
    \wh{\lambda} \defeq \argmax_\lambda L_{\text{max}}(\lambda)\ ,
\label{eq:hat_lambda}
\end{equation}
with approximate $100(1 - \alpha)$ percent confidence interval
\begin{equation}
    L_{\text{max}}(\wh{\lambda}) - L_{\text{max}}(\lambda) < \frac{1}{2} \chi^2_1(1 - \alpha)\ ,
\label{eq:hat_lambda_ci}
\end{equation}
where $\chi^2_1(1 - \alpha)$ is the upper $\alpha$-quantile of the $\chi^2$ distribution with $1$ degree of freedom \cite{Box1964}. For most applications, it suffices to perform the optimisation \eqref{eq:hat_lambda} over small values of $\lambda$ in order to preserve interpretability and avoid extreme distortions of the data. In fact, there are good theoretical reasons to restrict to $\lambda \in [0, 1]$: this covers the entire spectrum from exponential to linear growth, which is the relevant range for most applications, and also ensures finite long-run variance \cite{Pirjol2017}.

In the special case that the increments are assumed to follow an \emph{i.i.d.} Gaussian process, \eqref{eq:llf_max} simplifies considerably to
\begin{equation}
    L_{\text{max}}(\lambda) = -\frac{T-1}{2} \log{\wh{\sigma}^2(\lambda)}\ ,
\label{eq:llf_max_iid}
\end{equation}
where $\wh{\sigma}^2(\lambda)$ is the sample variance of $\Delta F(\bm{x}, \lambda)$. This formulation makes transparent the ultimate aim of the procedure, which is to locate the $\hat{\lambda}$ which minimises the variance of the \emph{increments} of the transformed process; in contrast, the standard Box-Cox procedure simply minimises the variance of the transformed observations themselves. For many applications, \eqref{eq:llf_max_iid} is a good starting point since calculations are instantaneous and the likelihood function often smooth. This case studies in Section~\ref{sec:case_study} assume this specification.

It is clear that that the above discussion can be straightforwardly extended to time series which require a combination of transformation followed by $n \in \{0, 1, 2, \ldots \}$ differences, denoted $\Delta^{(n)} F(\bm{x}, \lambda)$, to achieve statistical ergodicity.\footnote{Fractional differencing is typically associated with long-range dependence, under which standard ergodicity conditions fail \cite{Granger1980}; such processes are therefore outside the scope of the present framework.} In the special case $n = 0$, the ergodicity transformation reduces to the Box-Cox transformation. An fundamental example of this case is the \emph{Ornstein-Uhlenbeck} process, which is stationary and ergodic in levels, and thus requires no transformation \cite{Mardoukhi2020}. By contrast, the one-factor commodity model proposed by \citeA{Schwartz1997} evolves according to a geometric Ornstein-Uhlenbeck process, making the logarithm the appropriate ergodicity transformation. The case study in Section~\ref{sec:case_study_fred} documents time series that require $n = 2$ differences before ergodicity is attained.

Although Gaussian processes provide a simple yet flexible setting, many applied time series exhibit richer dependence structures; in such cases, ARMA and GARCH models furnish natural alternatives. The extension of the above procedure beyond Gaussian processes is conceptually straightforward, and existing results and algorithms for these broader model classes can be directly leveraged. The class of ARMA models for instance, which is of fundamental importance due to Wold's decomposition theorem \cite[Ch.~4.8]{Hamilton1994}, satisfies the following ergodic property: a covariance-stationary ARMA$(p, q)$ process with reciprocal roots inside the unit circle is ergodic for the first and second moments \cite[Ch.~7.2]{Hamilton1994}. Similar ergodicity results are available for the GARCH$(1, 1)$ model, which provides a natural starting point for applications in finance \cite{Francq2006, Hansen2005}. In each case, the problem reduces to finding the $\lambda$ which maximises the log likelihood of a candidate ARMA or GARCH model fitted to $\Delta^{(n)} F(\bm{x}, \lambda)$.

The final step of the proposed procedure is diagnostic analysis, aimed at confirming that the transformed process $\Delta^{(n)} F(\bm{x}, \lambda)$ is well-approximated by a stationary ergodic model from the chosen class. In the Gaussian setting, this reduces to verifying approximate normality, stationarity, and short memory. Normality can be assessed using the Shapiro–Wilk test and visually via a Q–Q plot; stationarity may be checked using both the KPSS and ADF tests for robustness; and short-memory behaviour can be evaluated by applying the Ljung–Box test for autocorrelation and by inspecting ACF/PACF plots to confirm that correlations decay sufficiently quickly. For ARMA and GARCH specifications, model-based diagnostics become appropriate: one may examine the whiteness of residuals, conduct Ljung–Box tests on standardised residuals, and in the GARCH case, additionally test for remaining ARCH effects to ensure volatility clustering has been adequately captured. Finally, the robustness of the estimated parameter $\wh{\lambda}$ can be probed by bootstrapping, i.e. recomputing \eqref{eq:hat_lambda} across sub-samples of the transformed increments.

\section{Case studies}
\label{sec:case_study}

This section presents two case studies. The first is a simulation study which employs geometric and arithmetic Brownian motion to contrast the ergodicity transformation with the Box-Cox transformation. The second case study explores the effect of data transformations on the factor analysis of a large macroeconomic database.

\subsection{Geometric \& arithmetic Brownian motion}
\label{sec:case_study_gbm}

\begin{figure}
    \centering
    \includegraphics[width=\linewidth]{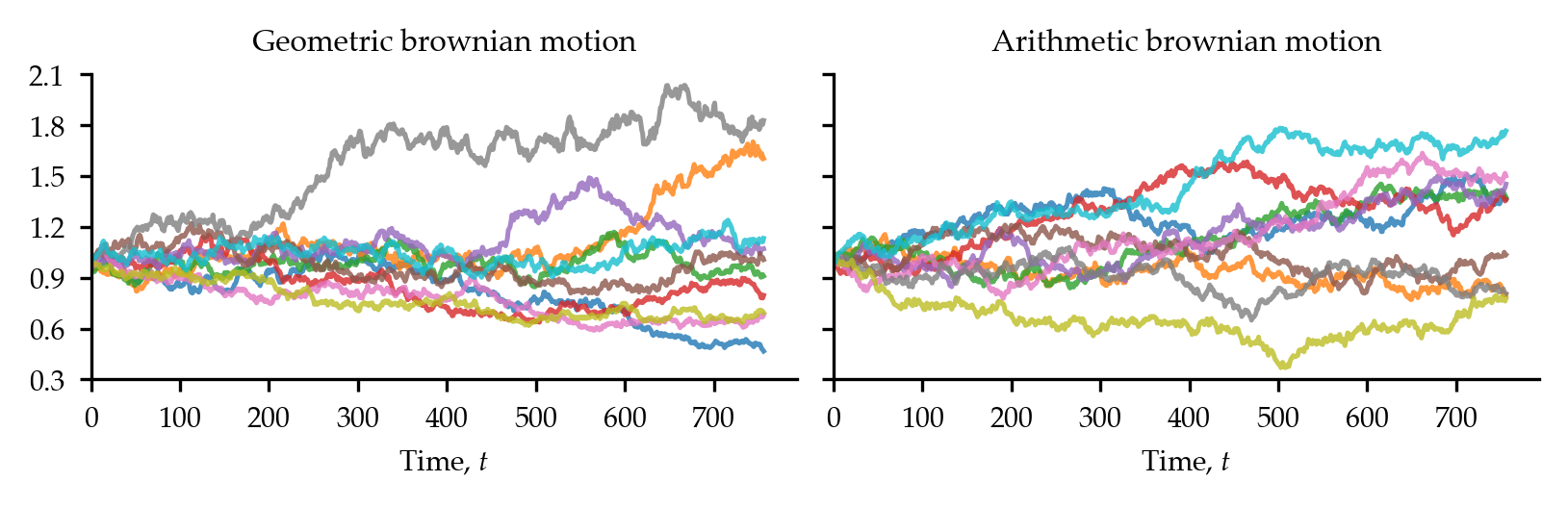}
    \caption{Ten sample trajectories each of geometric and arithmetic Brownian motion with drift $\mu =\qty{0.05}{\per\year}$, volatility $\sigma = \qty{0.2}{\per\year}$, and initial condition $X_0 = 1$, simulated over a period of 3 trading years (252 days per trading year).}
    \label{fig:gbm_paths}
\end{figure}

This case study investigates the extent to which the proposed method recovers known ergodicity transformations. Figure \ref{fig:gbm_paths} shows ten sample trajectories each of geometric and arithmetic Brownian motion (GBM and ABM respectively) with initial condition $X_0 = 1$, drift $\mu =\qty{0.05}{\per\year}$, and volatility $\sigma = \qty{0.2}{\per\year}$, simulated over a period of 3 trading years (252 days per trading year). We recall that an ABM follows
\begin{equation}
    X_t = X_0 + \mu t + \sigma W_t\ ,
\end{equation}
where $W_t$ is the Wiener process defined in \eqref{eq:wiener}.

\begin{figure}
    \centering
    \includegraphics[width=\linewidth]{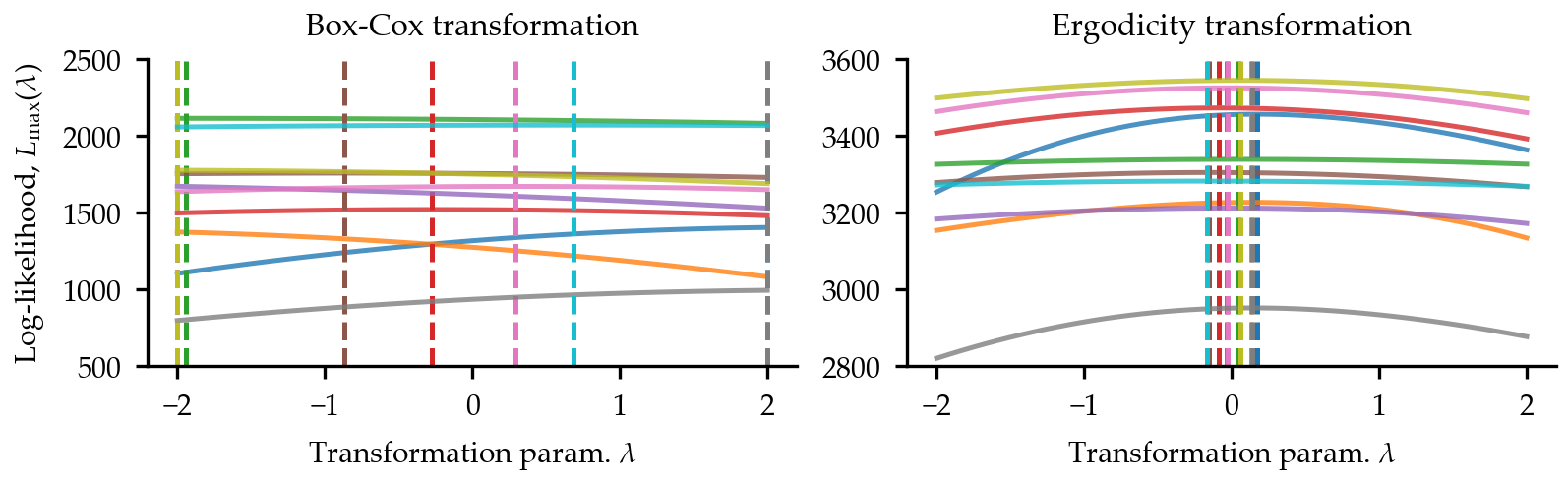}
    \caption{Left panel: the Box-Cox procedure applied to each GBM path from Figure~\ref{fig:gbm_paths}. For each path, the estimated transformation parameter $\wh{\lambda}$ is indicated by a dashed vertical line. Right panel: the profile log-likelihood $L_{\text{max}}(\lambda)$ defined in \eqref{eq:llf_max}, together with the corresponding estimate $\wh{\lambda}$ from \eqref{eq:hat_lambda}. The true transformation parameter is $\lambda = 0$.}
    \label{fig:gbm_llfs}
\end{figure}

\begin{figure}
    \centering
    \includegraphics[width=\linewidth]{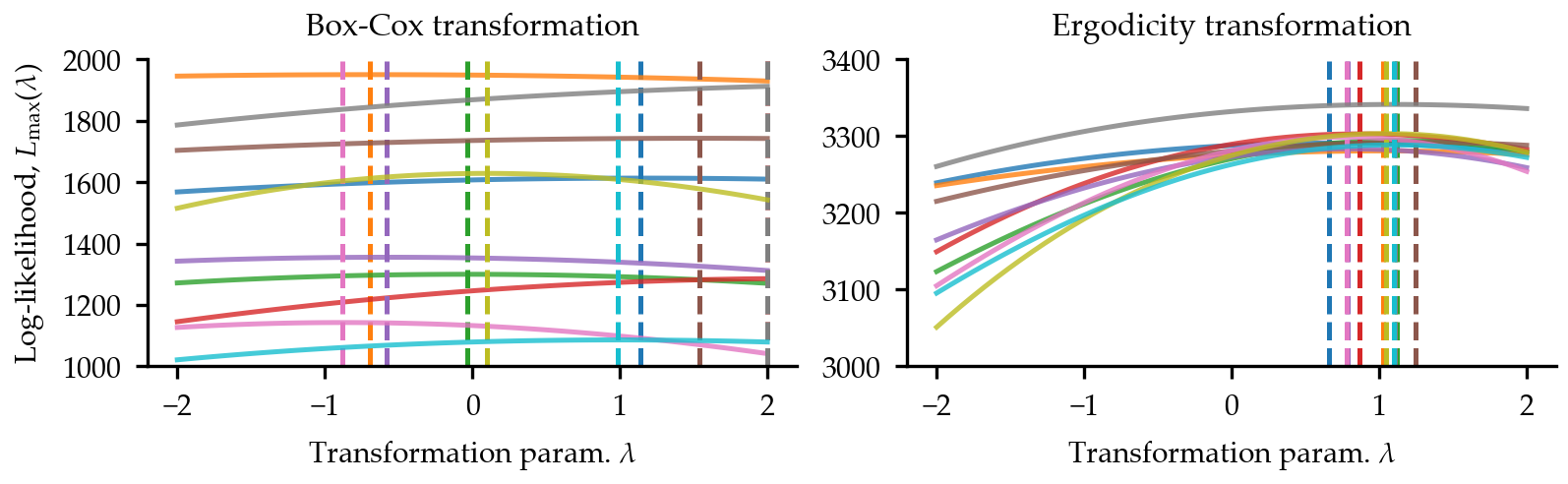}
    \caption{Same as Figure~\ref{fig:gbm_llfs}, but for the ABM paths from Figure~\ref{fig:gbm_paths}. The true transformation parameter is $\lambda = 1$.}
    \label{fig:abm_llfs}
\end{figure}

As discussed above, the correct transformation in the case of the GBM is the $\Delta\log$, which corresponds to $\lambda = 0$ for the power transformation in \eqref{eq:boxcox}. The right panel of Figure~\ref{fig:gbm_llfs} demonstrates how, for each GBM path, the profile log-likelihood function $L_{\text{max}}(\lambda)$ from \eqref{eq:llf_max} is indeed maximised at approximately $\lambda = 0$.\footnote{The confidence intervals are omitted to avoid visual clutter, but are easily computed from \eqref{eq:hat_lambda_ci}.} This is contrasted with the Box-Cox procedure applied to the same paths, shown in the left panel of Figure~\ref{fig:gbm_llfs}; it is clear that simply minimising the variance of the paths in levels does not consistently recover the underlying growth dynamic. Similarly, Figure \ref{fig:abm_llfs} demonstrates that the proposed ergodicity procedure reliably approximates the correct transformation $\lambda = 1$ for the ABM paths, whereas the Box-Cox transformation again produces diffuse results.

\begin{figure}
    \centering
    \includegraphics[width=\linewidth]{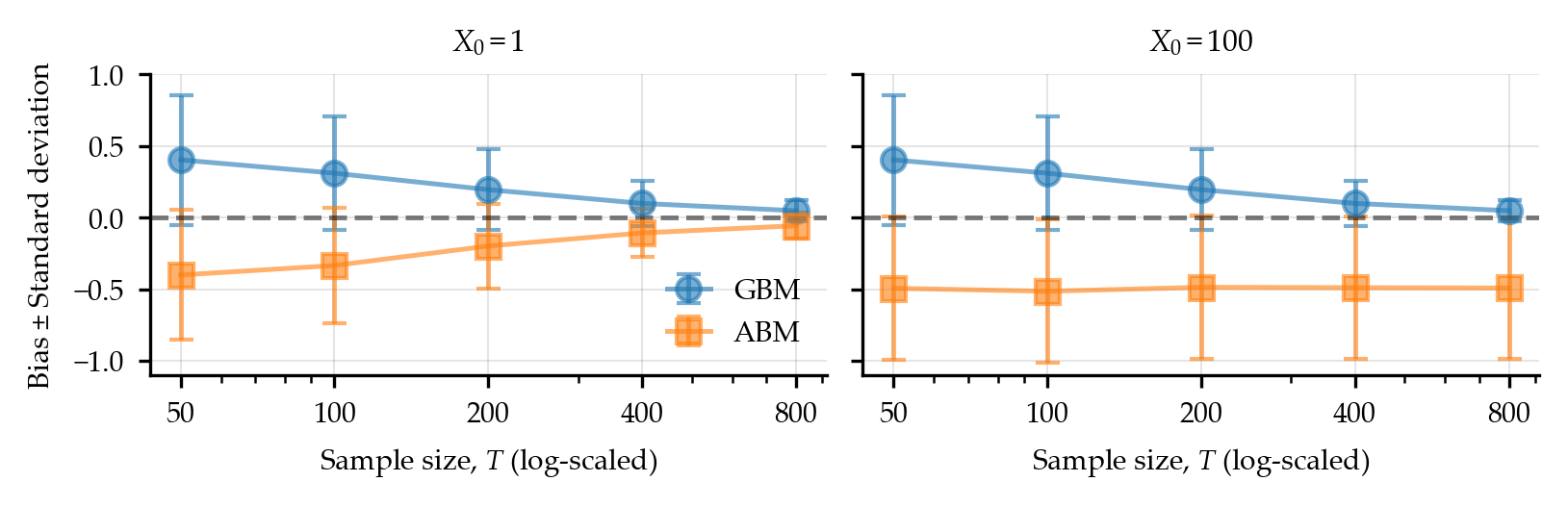}
    \caption{Effects of sample size on the accuracy of the proposed procedure. The bias and standard deviation of $\wh{\lambda}$ relative to the true values ($\lambda = 0$ for GBM; $\lambda = 1$ for ABM) are reported. Left panel: initial condition, $X_0 = 1$. Right panel: $X_0 = 100$.}
    \label{fig:lambda_hist}
\end{figure}

The accuracy and coverage probability of the proposed procedure are visually summarised in Figure~\ref{fig:lambda_hist}. For each sample size $T \in \{50, 100, 200, 400, 800 \}$, with $X_0$, $\mu$, and $\sigma$ fixed as above, one thousand paths each of GBM and ABM were simulated, $\wh{\lambda}$ estimated for each path, and the bias and standard deviation from the true value ($\lambda = 0$ for GBM; $\lambda = 1$ for ABM) recorded. The left panel shows that although some bias exists at small sample sizes, this is largely eliminated as the sample size increases. On the other hand, the right panel documents a limitation of the proposed procedure, which is related to the signal-to-noise ratio. In the example, if the level of the initial condition $X_0$ is increased by a factor of 100, the relative changes of the ABM become tiny compared to the overall level of the series; this flattens the profile log-likelihood curve, making identification of the theoretical value difficult. A similar issue is not observed for the GBM; the data-generating process here is such that the profile log-likelihood curve retains sufficient curvature to allow reliable identification of the true transformation parameter. This highlights that the performance of the procedure depends not only on sample size but also on the scale of the process relative to its stochastic variability, with additive models being more sensitive to low signal-to-noise regimes than their multiplicative counterparts.

More generally, when the data-generating process is unknown, the results suggest that inference on the transformation parameter should be accompanied by diagnostic assessments of identification strength, including the confidence interval of the estimate as given in \eqref{eq:hat_lambda_ci}, an examination of the curvature of the profile log-likelihood, and a test of whether the variance of increments is sufficiently large relative to the level of the series. A flat likelihood surface should be interpreted as evidence of weak signal rather than as support for any particular transformation.

\subsection{Dynamic factor analysis of a large macroeconomic database}
\label{sec:case_study_fred}

The second case study considers the effects of data transformations on a dynamic factor model of a large macroeconomic database. Dynamic factor analysis is predicated upon the assumption that the common dynamics of a large number of time series stem from a relatively small number of unobserved, or latent, factors, which in turn evolve over time \cite{Stock2016}. Once such factors have been extracted, they can be used for many purposes, including recession dating, forecasting, measuring uncertainty, and evaluating monetary policy \cite{McCracken2020}. This section performs a factor analysis on the \enquote{FRED-QD} database, maintained by the Federal Reserve of St Louis, consisting of 248 economic and financial quarterly series with broad coverage of the U.S.~economy; for interpretability, the series are classified into 14 groups \cite{McCracken2020}.\footnote{\label{footnote:groups}The groups are as follows: (1) National Income and Product Accounts (NIPA); (2) Industrial Production; (3) Employment and Unemployment; (4) Housing; (5) Inventories, Orders, and Sales; (6) Prices; (7) Earnings and Productivity; (8) Interest Rates; (9) Money and Credit; (10) Household Balance Sheets; (11) Exchange Rates; (12) Other; (13) Stock Markets; and (14) Non-Household Balance Sheets.}

For extracting latent factors, it is important that the considered series be stationary. For this purpose, FRED-QD provides a transformation code for each series, indicating whether it is stationary in levels, denoted $I(0)$, contains a unit root, denoted $I(1)$, or is integrated of order two, denoted $I(2)$. The transformation codes also give a recommendation as to whether the series should be treated in levels or log-levels before differencing. Although \citeA{McCracken2020} goes into some detail on the unit-root testing, the procedure underlying the levels vs.~log-levels decision is not described. Presumably, series exhibiting level-dependent variance are log-transformed to stabilise the variance, as is standard practice in this setting (cf. \citeNP{Hyndman2018}). This case study revisits this decision, and further examines how a principled choice of power transformation influences the subsequent factor analysis.

\begin{figure}
    \centering
    \includegraphics[width=\linewidth]{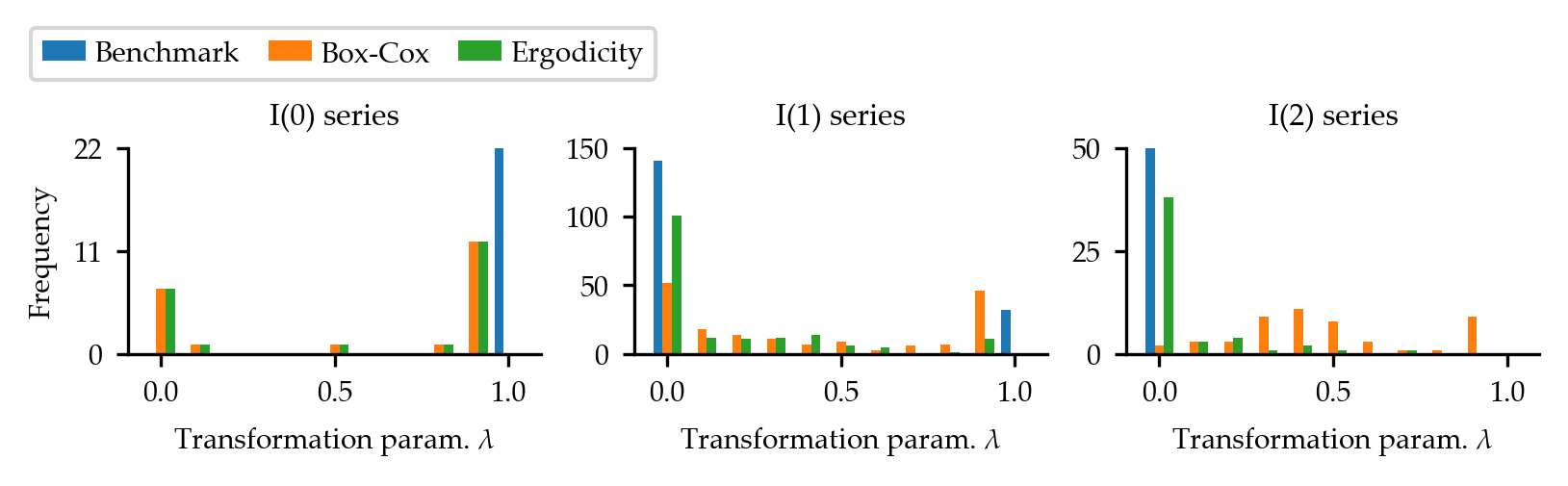}
    \caption{Distribution of estimated power transformation parameters $\wh{\lambda}$ across FRED-QD series, by order of integration, under the Box-Cox and ergodicity procedures. Benchmark FRED-QD transformations ($\lambda=1$ for levels, $\lambda=0$ for logs) are shown for comparison.}
    \label{fig:fred_hist}
\end{figure}

To this end, holding fixed the order of integration in the original dataset, Figure~\ref{fig:fred_hist} plots histograms of estimates for the power transformation parameter $\wh{\lambda}$ for all 248 series in FRED-QD, for both the Box-Cox and ergodicity transformation procedures. For comparison, the benchmark transformations provided by FRED-QD are also shown, with $\lambda = 1$ corresponding to no transformation and $\lambda = 0$ to a log transformation. First, for the $I(0)$ series (22 in total), the benchmark retains all series in levels; by contrast, the Box-Cox and ergodicity procedures, which coincide in this case, depart from this choice for approximately half of the series. For the $I(1)$ series (173 in total), the benchmark applies a log transformation to 141 series and no transformation to the remaining 32; by contrast, both the Box-Cox and ergodicity transformations each shift part of the mass of the estimated $\wh{\lambda}$ distribution into the interval $(0,1)$. Visually, the ergodicity transformation appears to strike a compromise between the benchmark and Box-Cox selections, with the latter moving substantial mass away from the log transformation. A similar pattern is observed, even more strongly, for the $I(2)$ series (50 in total), where the Box-Cox procedure almost entirely eschews the log transformation in favour of power transformations with $\wh{\lambda} \in (0,1)$.

\begin{figure}
    \centering
    \includegraphics[width=\linewidth]{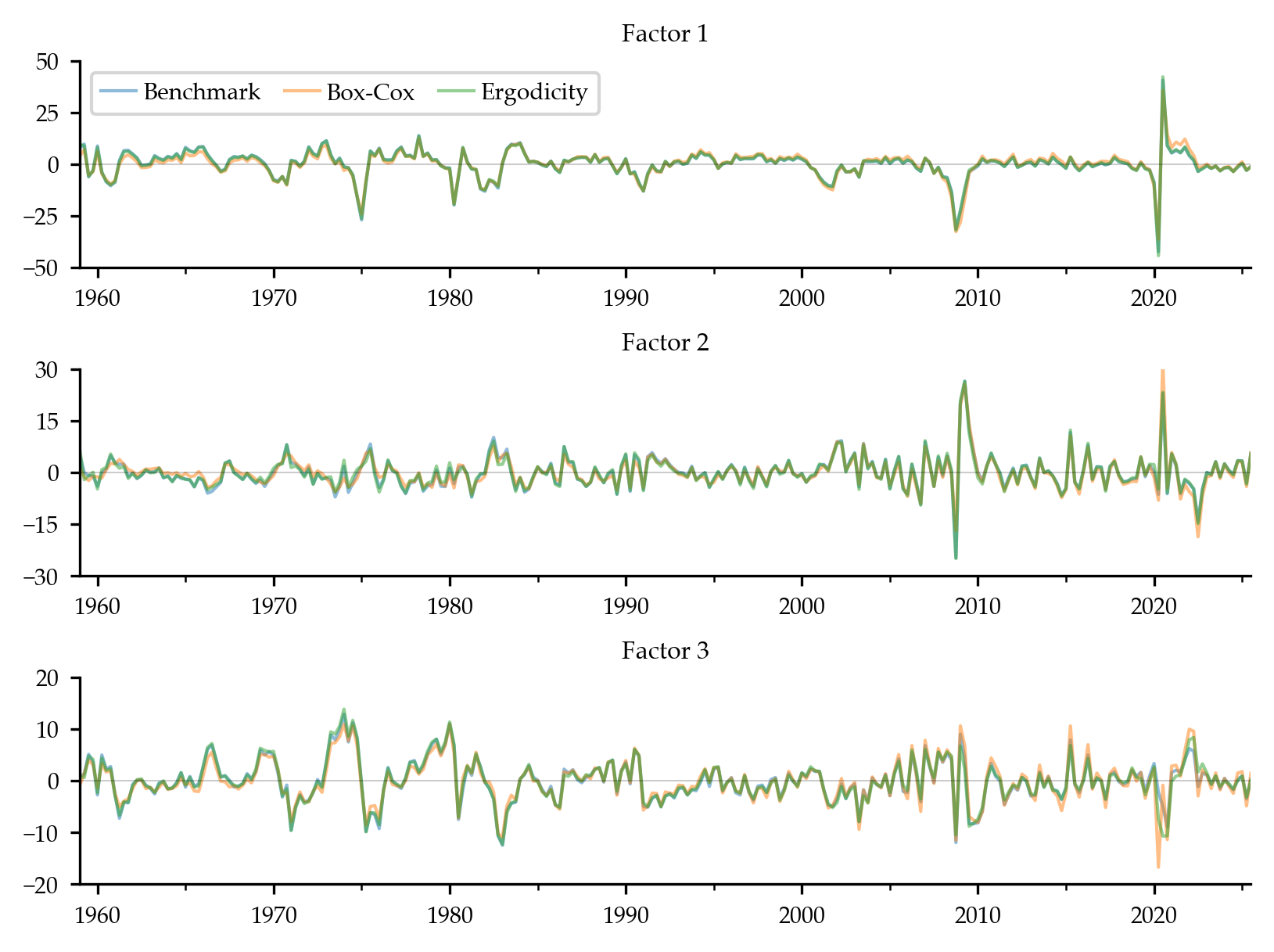}
    \caption{Visual comparison of the extracted factors under three specifications: benchmark transformations from FRED-QD, the Box-Cox transformation, and the ergodicity transformation.}
    \label{fig:fred_factors}
\end{figure}

\begin{figure}
    \centering
    \includegraphics[width=\linewidth]{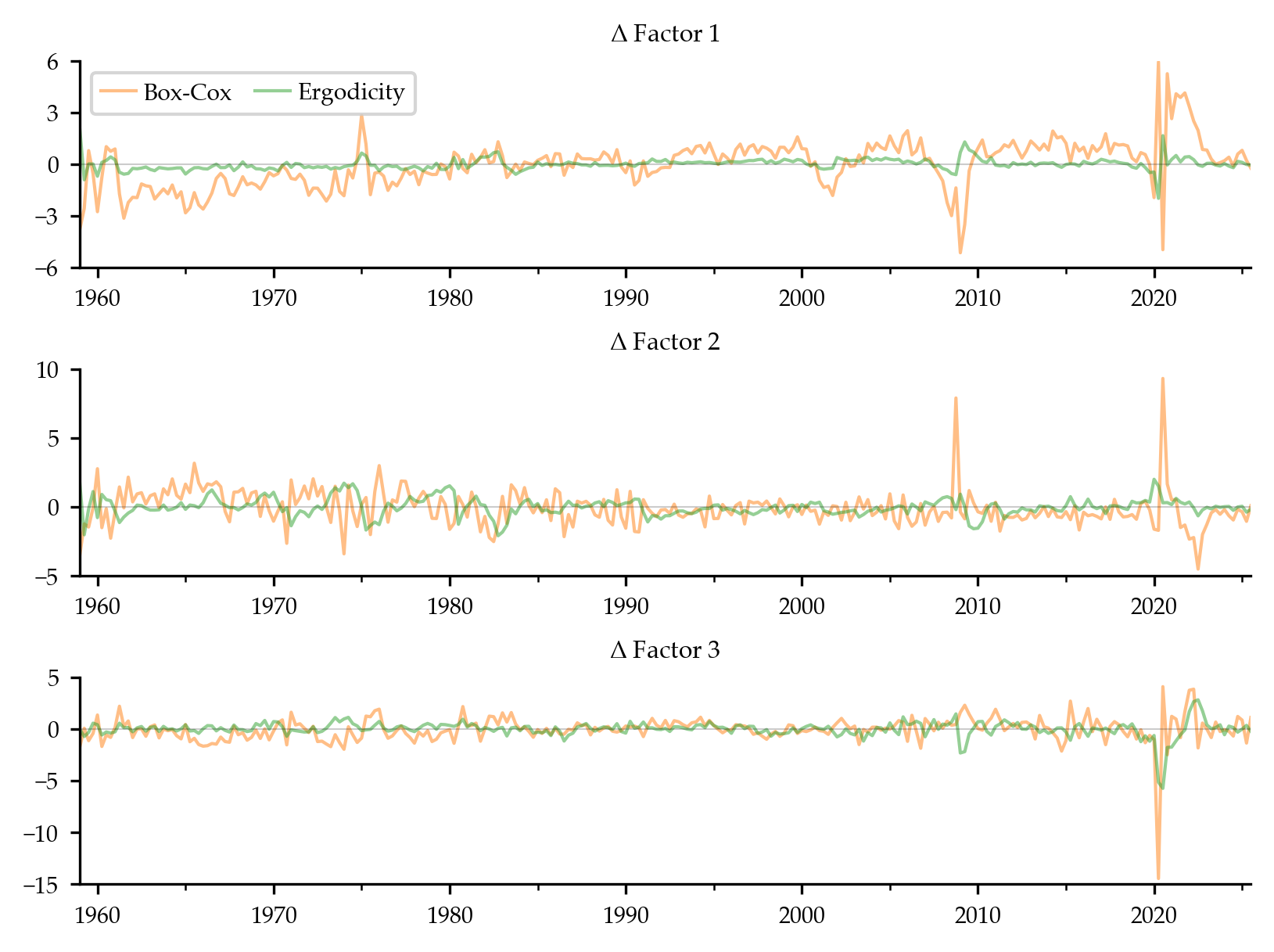}
    \caption{Difference to the benchmark factors for the Box-Cox and ergodicity specifications (cf.~Figure~\ref{fig:fred_factors}).}
    \label{fig:fred_factors_diff}
\end{figure}

Based on these estimated transformations, the expectation maximisation algorithm of \citeA{Banbura2012}, as implemented in the Python package \enquote{statsmodels} \cite{Fulton2022}, is then employed to fit a dynamic factor model under the three specifications (benchmark, Box-Cox, ergodicity). After some trial and error, we restrict attention to three global factors based on explanatory power (cf.~\citeNP{McCracken2020}). The extracted factors are displayed in Figure~\ref{fig:fred_factors}. They are visually similar across specifications and, under each transformation, jointly explain approximately \qty{34}{\percent} of the total variation across all series in the data set. Nevertheless, this apparent similarity masks systematic differences in how variation is allocated across factors and over time. These differences are more clearly revealed in Figure~\ref{fig:fred_factors_diff}, which plots the difference between the Box–Cox and ergodicity-based factors from their benchmark counterparts. First, it is clear that both specifications exhibit high-frequency deviations from the benchmark, with those under Box–Cox being more pronounced. These deviations are particularly notable during periods of heightened macroeconomic stress, reflecting the Box–Cox specification’s tendency to overweight extreme events. Relative to the benchmark, the Box–Cox factors display a low-frequency upward trend in Factor~1 and cyclical components in Factors~2 and 3, suggesting that variance-stabilising transformations can alter the representation of fundamental dynamics and introduce disagreement in the extracted latent factors. This motivates a closer examination of factor loadings and marginal explanatory power across specifications.

Therefore, we follow \citeA{McCracken2020} and regress the $i$-th series in the data set on each of the $k$ factors; for $k = 1, \ldots, 4$, this yields coefficients of determination $R_i^2(k)$ for each series $i$. Because the factors are orthogonal and organized in decreasing order of their respective eigenvalues, the incremental explanatory power of factor $k$ for series $i$ is $mR^2_i(k) = R^2_i(k) - R^2_i(k-1)$, $k = 2, 3, 4$, with $mR^2_i(1) = R^2_i(1)$. The average importance of factor $k$ is then given by $mR^2(k) = \tfrac{1}{N}\sum_{i=1}^N mR^2_i(k)$. Tables~\ref{tab:factor_1}--\ref{tab:factor_3} each list the top ten loadings per factor, as well as the overall factor importance $mR^2(k)$, with Table~\ref{tab:key} providing a descriptive key for the FRED-QD mnemonics. We discuss each factor in turn.

First, Table~\ref{tab:factor_1} shows that Factor~1 is highly robust across all three specifications and admits a clear economic interpretation as a broad \emph{real-activity factor}. In the benchmark model, it loads most strongly on private and non-farm employment (USPRIV, PAYEMS), manufacturing and industrial production (IPMANSICS, INDPRO, OUTMS), and hours worked (HOANBS, HOABS), indicating that it primarily captures aggregate labour-market conditions and goods-producing activity. The Box-Cox specification yields a very similar composition, with employment, hours, and production measures again dominating, alongside closely related indicators such as the unemployment rate (UNRATE) and the help-wanted index (HWIURATIOx), reflecting modest re-ranking rather than substantive change. Under the ergodicity transformation, the factor remains anchored in employment and manufacturing output (USPRIV, PAYEMS, IPMANSICS, INDPRO), with hours worked and goods-sector employment continuing to feature prominently. The stability of both the contributing series and their marginal explanatory power across transformations supports the interpretation of Factor~1 as a persistent and economically meaningful measure of aggregate real activity.

\begin{table}
\centering
\scriptsize
    \begin{tabular}{lcclcclcc}
        \toprule
        \multicolumn{3}{c}{Benchmark ($mR^2(1) = 0.2033$)}  & \multicolumn{3}{c}{Box-Cox ($mR^2(1) = 0.1974$)}  & \multicolumn{3}{c}{Ergodicity ($mR^2(1) = 0.2075$)} \\
        \cmidrule(r){1-3} \cmidrule(lr){4-6} \cmidrule(l){7-9}
        Series & $mR^2$ & Group & Series & $mR^2$ & Group & Series & $mR^2$ & Group \\
        \midrule
        USPRIV & 0.8591 & 3 & USPRIV & 0.8139 & 3 & USPRIV & 0.8493 & 3 \\
        PAYEMS & 0.8305 & 3 & HOANBS & 0.8111 & 3 & IPMANSICS & 0.8425 & 2 \\
        IPMANSICS & 0.8293 & 2 & HWIURATIOx & 0.7989 & 3 & PAYEMS & 0.8188 & 3 \\
        HOANBS & 0.8176 & 3 & USGOOD & 0.7931 & 3 & HOANBS & 0.8185 & 3 \\
        INDPRO & 0.8156 & 2 & HOABS & 0.7883 & 3 & INDPRO & 0.7991 & 2 \\
        MANEMP & 0.7821 & 3 & UNRATE & 0.7880 & 3 & USGOOD & 0.7920 & 3 \\
        HOABS & 0.7821 & 3 & PAYEMS & 0.7839 & 3 & HOABS & 0.7846 & 3 \\
        USGOOD & 0.7722 & 3 & LNS13023621 & 0.7793 & 3 & OUTMS & 0.7753 & 1 \\
        OUTMS & 0.7664 & 1 & INDPRO & 0.7689 & 2 & HOAMS & 0.7501 & 3 \\
        DMANEMP & 0.7606 & 3 & IPMANSICS & 0.7653 & 2 & MANEMP & 0.7451 & 3 \\
        \bottomrule
    \end{tabular}
\caption{Top ten series contributing to Factor 1 across three specifications: benchmark (from FRED-QD), Box-Cox, and ergodicity-transformed data. Series are ranked by marginal $mR_i^2$ for each factor, with the corresponding group and contribution indicated. The overall factor importance $mR^2(1)$ is reported for each specification. Table~\ref{tab:key} provides a key for the FRED-QD mnemonics.}
\label{tab:factor_1}
\end{table}

Table~\ref{tab:factor_2} indicates that Factor~2 is consistently identified as a \emph{price and inflation factor} across all three specifications. In the benchmark model, the factor loads almost exclusively on consumer and producer price measures, with headline and core CPI components (CPITRNSL, CUSR0000SAC, CUSR0000SA0L2), consumption deflators (PCECTPI), and durable and nondurable goods price indices (DGOERG3Q086SBEA and similar) accounting for the bulk of its explanatory power. The Box-Cox specification preserves this interpretation but induces a modest reshuffling within the price block, placing relatively greater weight on alternative CPI definitions and wholesale price indices (e.g.~WPSID61), and is associated with a slightly higher overall importance, as reflected in $mR^2(2)$. Under the ergodicity transformation, the factor closely mirrors the benchmark both in composition and ranking, remaining dominated by CPI subcomponents and consumption deflators, with marginally higher contributions for several headline inflation series. The strong overlap in series identities and marginal explanatory power across specifications supports the interpretation of Factor~2 as a stable common inflation component, largely invariant to the transformation applied and reflecting differences in scaling rather than substantive changes in the underlying economic signal.

\begin{table}
\centering
\scriptsize
\setlength\tabcolsep{0pt}
    \begin{tabular*}{\linewidth}{@{\extracolsep{\fill}} lcclcclcc}
    \toprule
    \multicolumn{3}{c}{Benchmark ($mR^2(2) = 0.0747$)}  & \multicolumn{3}{c}{Box-Cox ($mR^2(2) = 0.0836$)}  & \multicolumn{3}{c}{Ergodicity ($mR^2(2) = 0.0727$)} \\
    \cmidrule{1-3} \cmidrule{4-6} \cmidrule{7-9}
    Series & $mR^2$ & Group & Series & $mR^2$ & Group & Series & $mR^2$ & Group \\
    \midrule
    CPITRNSL & 0.6128 & 6 & CUSR0000SA0L2 & 0.5620 & 6 & CPITRNSL & 0.6545 & 6 \\
    CUSR0000SAC & 0.6008 & 6 & CUSR0000SAC & 0.5577 & 6 & CUSR0000SAC & 0.6511 & 6 \\
    CUSR0000SA0L2 & 0.5779 & 6 & DGOERG3Q086SBEA & 0.5300 & 6 & CUSR0000SA0L2 & 0.6275 & 6 \\
    DGDSRG3Q086SBEA & 0.5720 & 6 & CPITRNSL & 0.5235 & 6 & DGDSRG3Q086SBEA & 0.6241 & 6 \\
    DGOERG3Q086SBEA & 0.5632 & 6 & WPSID61 & 0.5099 & 6 & DGOERG3Q086SBEA & 0.5987 & 6 \\
    DNDGRG3Q086SBEA & 0.5498 & 6 & DGDSRG3Q086SBEA & 0.5089 & 6 & DNDGRG3Q086SBEA & 0.5954 & 6 \\
    PPIIDC & 0.4986 & 6 & PCECTPI & 0.4999 & 6 & PCECTPI & 0.5491 & 6 \\
    PCECTPI & 0.4949 & 6 & CPIULFSL & 0.4949 & 6 & PPIIDC & 0.5358 & 6 \\
    CUSR0000SA0L5 & 0.4767 & 6 & CUSR0000SA0L5 & 0.4855 & 6 & CUSR0000SA0L5 & 0.5291 & 6 \\
    PPIACO & 0.4702 & 6 & CPIAUCSL & 0.4819 & 6 & CPIAUCSL & 0.5224 & 6 \\
    \bottomrule
    \end{tabular*}
\caption{Same as Table~\ref{tab:factor_1}, but for Factor 2.}
\label{tab:factor_2}
\end{table}

Table~\ref{tab:factor_3} highlights a marked divergence in the economic coherence of Factor~3 across transformations. In the benchmark specification, the factor exhibits a mixed composition, combining investment and inventory measures (PRFIx, BUSINVx), interest-rate spreads (T5YFFM, AAAFFM), and several consumer price indices. This overlap with price-level series already prominent in Factor~2 weakens the interpretability of Factor~3, suggesting residual inflationary contamination rather than a distinct economic dimension. The Box-Cox specification further accentuates this ambiguity: while real money balances (M2REAL) enter prominently, the factor is otherwise dominated by CPI and PCE deflators, leaving it difficult to disentangle monetary dynamics from the inflation block identified in the preceding factor. By contrast, the ergodicity transformation yields a substantially sharper and more economically coherent factor. Notably, none of the top-loading series belong to the price-index group, eliminating the overlap with Factor~2 that characterises the other specifications. Instead, Factor~3 under ergodicity is organised around real money aggregates (M2REAL, M1REAL), interest-rate spreads (AAAFFM, T5YFFM), and real activity in investment, inventories, and housing (PRFIx, BUSINVx, PERMIT, HOUST). This composition admits a clearer interpretation as a \emph{financial and monetary conditions factor}, capturing liquidity, financing conditions, and interest-rate sensitivity rather than price-level movements. The absence of inflation measures and the tighter clustering of economically related series underscore the extent to which the ergodicity transformation isolates a genuinely distinct source of macroeconomic variation, providing the clearest separation between nominal and real-financial dynamics in the factor structure.

\begin{table}
\centering
\scriptsize
    \begin{tabular}{lcclcclcc}
    \toprule
    \multicolumn{3}{c}{Benchmark ($mR^2(3) = 0.0584$)}  & \multicolumn{3}{c}{Box-Cox ($mR^2(3) = 0.0605$)}  & \multicolumn{3}{c}{Ergodicity ($mR^2(3) = 0.0618$)} \\
    \cmidrule(r){1-3} \cmidrule(lr){4-6} \cmidrule(l){7-9}
    Series & $mR^2$ & Group & Series & $mR^2$ & Group & Series & $mR^2$ & Group \\
    \midrule
    PRFIx & 0.2983 & 1 & M2REAL & 0.3250 & 9 & PRFIx & 0.3033 & 1 \\
    T5YFFM & 0.2775 & 8 & CPIAUCSL & 0.3239 & 6 & BUSINVx & 0.2912 & 5 \\
    AAAFFM & 0.2667 & 8 & CUSR0000SA0L5 & 0.3111 & 6 & M2REAL & 0.2895 & 9 \\
    CPIAUCSL & 0.2379 & 6 & CPIULFSL & 0.3104 & 6 & AAAFFM & 0.2867 & 8 \\
    BUSINVx & 0.2285 & 5 & PCECTPI & 0.2984 & 6 & T5YFFM & 0.2846 & 8 \\
    CUSR0000SA0L5 & 0.2264 & 6 & CPITRNSL & 0.2867 & 6 & PERMIT & 0.2541 & 4 \\
    PERMITS & 0.2199 & 4 & CUSR0000SA0L2 & 0.2810 & 6 & PERMITS & 0.2457 & 4 \\
    PCECTPI & 0.2178 & 6 & DGDSRG3Q086SBEA & 0.2722 & 6 & TCU & 0.2424 & 2 \\
    CPIULFSL & 0.2123 & 6 & DNDGRG3Q086SBEA & 0.2645 & 6 & HOUST & 0.2416 & 4 \\
    DGDSRG3Q086SBEA & 0.2118 & 6 & CUSR0000SAC & 0.2537 & 6 & M1REAL & 0.2260 & 9 \\
    \bottomrule
    \end{tabular}
\caption{Same as Table~\ref{tab:factor_1}, but for Factor 3.}
\label{tab:factor_3}
\end{table}

\begin{table}
\centering
\scriptsize
    \begin{tabularx}{\linewidth}{lcX}
    \toprule
    FRED-QD mnemonic & Group & Description\\
    \midrule
    AAAFFM & 8 & Moody's Seasoned AAA Corporate Bond Minus Federal Funds Rate \\
    BUSINVx & 5 & Total Business Inventories (Millions of Dollars) \\
    CPIAUCSL & 6 & Consumer Price Index for All Urban Consumers: All Items (Index 1982-84=100) \\
    CPITRNSL & 6 & Consumer Price Index for All Urban Consumers: Transportation (Index 1982-84=100) \\
    CPIULFSL & 6 & Consumer Price Index for All Urban Consumers: All Items Less Food (Index 1982-84=100) \\
    CUSR0000SA0L2 & 6 & Consumer Price Index for All Urban Consumers: All items less shelter (Index 1982-84=100) \\
    CUSR0000SA0L5 & 6 & Consumer Price Index for All Urban Consumers: All items less medical care (Index 1982-84=100) \\
    CUSR0000SAC & 6 & Consumer Price Index for All Urban Consumers: Commodities (Index 1982-84=100) \\
    DGDSRG3Q086SBEA & 6 & Personal consumption expenditures: Goods (chain-type price index) \\
    DGOERG3Q086SBEA & 6 & Personal consumption expenditures: Nondurable goods: Gasoline and other energy goods (chain-type price index) \\
    DMANEMP & 3 & All Employees: Durable goods (Thousands of Persons) \\
    DNDGRG3Q086SBEA & 6 & Personal consumption expenditures: Nondurable goods (chain-type price index) \\
    HOABS & 3 & Business Sector: Hours of All Persons (Index 2017=100) \\
    HOAMS & 3 & Manufacturing Sector: Hours of All Persons (Index 2017=100) \\
    HOANBS & 3 & Nonfarm Business Sector: Hours of All Persons (Index 2017=100) \\
    HOUST & 4 & Housing Starts: Total: New Privately Owned Housing Units Started (Thousands of Units) \\
    HWIURATIOx & 3 & Ratio of Help Wanted/No. Unemployed \\
    INDPRO & 2 & Industrial Production Index (Index 2017=100) \\
    IPMANSICS & 2 & Industrial Production: Manufacturing (SIC) (Index 2017=100) \\
    LNS13023621 & 3 & Unemployment Level - Job Losers (Thousands of Persons) \\
    M1REAL & 9 & Real M1 Money Stock (Billions of 1982-84 Dollars), deflated by CPI \\
    M2REAL & 9 & Real M2 Money Stock (Billions of 1982-84 Dollars), deflated by CPI \\
    MANEMP & 3 & All Employees: Manufacturing (Thousands of Persons) \\
    OUTMS & 1 & Manufacturing Sector: Real Output (Index 2017=100) \\
    PAYEMS & 3 & All Employees: Total nonfarm (Thousands of Persons) \\
    PCECTPI & 6 & Personal Consumption Expenditures: Chain-type Price Index (Index 2017=100) \\
    PERMIT & 4 & New Private Housing Units Authorized by Building Permits (Thousands of Units) \\
    PERMITS & 4 & New Private Housing Units Authorized by Building Permits in the South Census Region (Thousands, SAAR) \\
    PPIACO & 6 & Producer Price Index for All Commodities (Index 1982=100) \\
    PPIIDC & 6 & Producer Price Index by Commodity Industrial Commodities (Index 1982=100) \\
    PRFIx & 1 & Real private fixed investment: Residential (Billions of Chained 2017 Dollars), deflated using its own price index \\
    T5YFFM & 8 & 5-Year Treasury Constant Maturity Minus Federal Funds Rate \\
    TCU & 2 & Capacity Utilization: Total Industry (Percent of Capacity) \\
    UNRATE & 3 & Civilian Unemployment Rate (Percent) \\
    USGOOD & 3 & All Employees: Goods-Producing Industries (Thousands of Persons) \\
    USPRIV & 3 & All Employees: Total Private Industries (Thousands of Persons) \\
    WPSID61 & 6 & Producer Price Index by Commodity Intermediate Materials: Supplies \& Components (Index 1982=100) \\
    \bottomrule
    \end{tabularx}
\caption{Interpretive key for the FRED-QD mnemonics in Tables~\ref{tab:factor_1}--\ref{tab:factor_3}.}
\label{tab:key}
\end{table}

To assess whether these interpretative gains translate into forecasting improvements, a simple exploratory exercise is conducted. Using a rolling expanding window, one-step-ahead forecasts are generated under the three specifications for the final 48 quarterly observations in the sample, ending in September 2025. Forecast accuracy is evaluated using the mean absolute scaled error (MASE) of \citeA{Hyndman2006}, which measures performance relative to a naive benchmark. Table~\ref{tab:fred_forecast} reports mean and median MASE for the full panel and by group; values below one indicate outperformance of a random walk. The table also reports, for each group, the fraction of series whose median MASE is significantly below one at the \qty{5}{\percent} level, based on a Wilcoxon signed-rank test with Benjamini–Hochberg false discovery rate control.

\begin{table}
\centering
\scriptsize
\setlength\tabcolsep{0pt}
    \begin{tabular*}{\linewidth}{@{\extracolsep{\fill}} ccccccccccc}
        \toprule
         && \multicolumn{3}{c}{Mean MASE} & \multicolumn{3}{c}{Median MASE} & \multicolumn{3}{c}{Fraction median MASE < 1} \\
         \cmidrule{3-5} \cmidrule{6-8} \cmidrule{9-11}
         Group & \# Series & Benchmark & Box-Cox & Ergodicity & Benchmark & Box-Cox & Ergodicity & Benchmark & Box-Cox & Ergodicity \\
         \midrule
         All & 245 & 1.1433 & 1.3070 & 1.1565 & \cellcolor{mygreen!30}0.5779 & 0.6613 & 0.6076 & \cellcolor{mygreen!30}0.70 & 0.47 & 0.67 \\
         
         \midrule

         1   & 23 & 1.0032 & 1.2147 & 1.0555 & \cellcolor{mygreen!30}0.4245 & 0.5523 & 0.4580 & 0.96 & 0.83 & \cellcolor{mygreen!30}1.00 \\
        
         2   & 16 & 1.1639 & 1.4913 & 1.2084 & \cellcolor{mygreen!30}0.6201 & 0.7698 & 0.6619 & \cellcolor{mygreen!30}0.62 & 0.25 & 0.56 \\
        
         3   & 50 & 1.8168 & 1.9999 & 1.7259 & \cellcolor{mygreen!30}0.6046 & 0.6768 & 0.6552 & \cellcolor{mygreen!30}0.66 & 0.46 & 0.58 \\
        
         4   & 14 & 0.8049 & \cellcolor{mygreen!30}0.7895 & 0.7979 & 0.5076 & \cellcolor{mygreen!30}0.4836 & 0.4956 & 0.79 & 0.79 & 0.79 \\
        
         5   & 9  & 1.0349 & 1.2349 & 1.0829 & 0.6635 & 0.7119 & \cellcolor{mygreen!30}0.6512 & 0.67 & 0.33 & 0.67 \\
        
         6   & 48 & \cellcolor{mygreen!30}0.8077 & 1.0610 & 0.8342 & \cellcolor{mygreen!30}0.5599 & 0.7668 & 0.5701 & 0.81 & 0.38 & \cellcolor{mygreen!30}0.83 \\
        
         7   & 14 & \cellcolor{mygreen!30}0.8288 & 0.9864 & 0.8834 & \cellcolor{mygreen!30}0.4842 & 0.6333 & 0.5258 & 0.93 & 0.57 & 0.93 \\
        
         8   & 20 & \cellcolor{mygreen!30}0.7605 & 1.0338 & 0.9879 & \cellcolor{mygreen!30}0.5690 & 0.6705 & 0.6546 & \cellcolor{mygreen!30}0.70 & 0.40 & 0.40 \\
        
         9   & 15 & 1.2558 & 1.4013 & 1.4830 & \cellcolor{mygreen!30}0.6524 & 0.6657 & 0.7657 & \cellcolor{mygreen!30}0.60 & 0.33 & 0.40 \\
        
         10  & 9  & 1.2865 & 1.2623 & 1.1971 & \cellcolor{mygreen!30}0.6258 & 0.6387 & 0.6551 & 0.11 & 0.11 & \cellcolor{mygreen!30}0.22 \\
        
         11  & 6  & 0.7629 & 0.7652 & \cellcolor{mygreen!30}0.7111 & 0.5333 & 0.5457 & \cellcolor{mygreen!30}0.4911 & 0.83 & 0.83 & 0.83 \\
        
         12  & 2  & 1.2450 & 1.0963 & 1.1082 & 0.9264 & 0.9434 & \cellcolor{mygreen!30}0.8991 & 0.00 & 0.50 & 0.50 \\
        
         13  & 6  & \cellcolor{mygreen!30}0.6741 & 0.8113 & 0.7391 & \cellcolor{mygreen!30}0.4625 & 0.5541 & 0.5157 & 0.83 & 0.83 & 0.83 \\
        
         14  & 13 & 1.5296 & 1.2930 & 1.2633 & 0.9141 & 0.8069 & \cellcolor{mygreen!30}0.7670 & \cellcolor{mygreen!30}0.38 & 0.15 & 0.15 \\
         \bottomrule
    \end{tabular*}
\caption{Mean and median MASE, and the fraction of series for which a model significantly outperforms the benchmark using a Wilcoxon test with false discovery rate control. Cells are highlighted when a model is the unique winner within a group (mean/median strictly below one; fraction strictly highest). A description of the 14 groups is found in Footnote~\ref{footnote:groups}.}
\label{tab:fred_forecast}
\end{table}

Several clear patterns emerge from Table \ref{tab:fred_forecast}. First, mean and median MASE convey systematically different information. Mean MASE frequently exceeds one, reflecting the disproportionate influence of a small subset of highly volatile or poorly forecastable series, whereas median MASE is below one in nearly all groups and specifications, indicating that the typical series is forecast more accurately than a random walk. This divergence is most pronounced in Industrial Production (2), Employment and Unemployment (3), and Money and Credit (9), where median performance remains favourable despite elevated mean errors. These categories therefore appear characterised by substantial cross-sectional heterogeneity, with a minority of series driving poor average performance rather than broad-based forecast failure.

Across specifications, the Box-Cox transformation underperforms consistently, in line with existing results in the macro-forecasting literature \cite{Nelson1979,Proietti2013}. It has the highest mean and median MASE in the aggregate and is rarely the unique winner within groups. Its weaker performance is particularly evident in Prices (6), Earnings and Productivity (7), Interest Rates (8), and Stock Markets (13), where both central tendency and the fraction of significant improvements are dominated by the alternatives. Box-Cox also records the lowest fraction of series with statistically significant median gains in most groups, reinforcing the impression that it degrades short-horizon forecast accuracy.

The ergodicity specification performs more competitively than either alternative. Although it does not dominate the benchmark in terms of aggregate mean MASE, it consistently improves upon Box–Cox and frequently emerges as the strongest or co-strongest specification across distributional metrics. In particular, for Inventories, Orders, and Sales (5), Money and Credit (9), Household Balance Sheets (10), and Non-Household Balance Sheets (14), ergodicity achieves either the lowest median MASE or the highest fraction of series with statistically significant gains. These improvements are especially pronounced for balance-sheet and credit aggregates, consistent with the interpretation that ergodicity-based transformations better respect the underlying accumulation and stock–flow dynamics of such variables. More generally, ergodicity displays a favourable robustness profile: even in groups where it is not the top performer, it rarely underperforms the benchmark by a meaningful margin, indicating stable and reliable forecasting performance across heterogeneous series.

\section{Conclusion}
\label{sec:discuss}

This article proposed a likelihood-based approach for estimating ergodicity transformations. The method is general, and can be readily integrated with standard time-series models, including Gaussian processes, ARMA, and GARCH. Through case studies, we demonstrated the method's ability to recover known ergodicity transformations and to enhance both the interpretability of latent factors and the accuracy of forecasting in economic and financial applications. The results suggest that incorporating ergodicity transformations can provide meaningful gains over conventional Box-Cox or naive approaches, without requiring model-specific modifications.

Future work might focus on a tighter integration of the method with ARIMA and GARCH models. In particular, we envision jointly estimating the power transformation alongside model-specific parameters, rather than following the current sequential approach of applying a Box-Cox transformation first and then fitting the model. Achieving this will require careful treatment of the interaction between non-linear transformations and unit-root testing (cf.~\citeNP{Franses1998}), ensuring that model estimation remains robust while capturing the benefits of ergodicity-based adjustments. Additionally, the framework could be extended to incorporate robust statistics, for example by minimizing variance or other dispersion measures with respect to robust estimators (cf.~\citeNP{Raymaekers2021}), which would improve resilience to outliers and heavy-tailed distributions common in applied work.

\bibliographystyle{apacite}
\bibliography{biblio}

\end{document}